\documentclass[12pt,]{article}
 \usepackage{graphicx}
 \usepackage[cp1251]{inputenc}
 \usepackage{epstopdf}
 \usepackage{rotating}
\begin{document}
\noindent {\small\it Astrophysics, V. 65, Issue 4, November 2022}

 \bigskip
 \bigskip
 \bigskip
 \centerline {\bf THE MYSTERIOUS RADCLIFFE WAVE}
 \bigskip
 \bigskip
 \centerline {V. V. Bobylev$^{1}$, A. T. Bajkova$^{1}$ and Yu. N. Mishurov$^{2}$}
 \bigskip
 {\small
\centerline{\it $^1$Central (Pulkovo) Astronomical Observatory of RAS, 65/1, Pulkovskoye Ch., 196140, St.-Petersburg}

\centerline{\it $^2$Southern Federal University, Rostov-on-Don}}

 \bigskip
 \bigskip
 \bigskip
{\bf Abstract.}
The review is devoted to the Radcliffe Wave recently discovered by Alves et
al. from the analysis of molecular clouds. These authors singled out a narrow chain
of molecular clouds, elongated almost in one line, located at an inclination of
about 30$^o$ to the galactic axis y. The Radcliffe Wave itself describes damped vertical
oscillations of molecular clouds with a maximum oscillation amplitude of about
160 pc and a characteristic wavelength of about 2.5 kpc. To date, the presence of
the Radcliffe Wave has been confirmed in the vertical distribution of a) interstellar
dust, b) sources of maser radiation and radio stars, which are very young stars
and protostars closely associated with molecular clouds, c) low-mass stars of the
T Tau type, d) more massive OB stars and e) young open clusters of stars. The
Radcliffe Wave is also traced in the vertical velocities of young stars. Most of the
considered results of the analysis of the vertical velocities of various young stars
show that the oscillations of the vertical positions and vertical velocities of stars
in the Radcliffe Wave occur synchronously. The nature of the Radcliffe Wave is
completely unclear. The majority of researchers associate its occurrence with the
assumption of an external gravitational impact on the galactic disk of a striker
such as a dwarf satellite galaxy of the Milky Way.

 \bigskip
 \bigskip
 \bigskip


\newpage
\section{Discovery of the Radcliffe Wave}
Near the Sun, the Radcliffe wave is known to propagate approximately along the Local Arm (Orion Arm). It was first discovered by Alves et al.~[1] from an analysis of the distribution of molecular clouds. The original authors of this research team are from the Radcliffe Institute for Advanced Study in Cambridge, Massachusetts. Therefore, they named the wave in honor of their native institute.

Alves et al. [1] identified a narrow chain of molecular clouds, elongated almost in one line, at an inclination of about $30^\circ$ to the galactic axis $y$. The wave itself is observed in the vertical coordinates $z$ of the clouds. That is, the structure is three-dimensional. According to Alves et al.~[1], the wave is damped and the maximum value of the amplitude is observed in the immediate vicinity of the Sun, where the Gould Belt is located~[2]. As noted by Alves et al.~[1], the Radcliffe wave should play an important role in understanding the reason for the formation of the Gould Belt. In particular, they believe that the presence of vertical disturbances closes Blaauw's hypothesis~[3] about the hypernova explosion.

The detection of the Radcliffe wave became possible due to the work of Zucker et al. [4, 5] who estimated the distances to molecular clouds in the Local Arm region located at heliocentric distances from 150~pc to 2.5~kpc. These authors developed a method that combines photometric data with trigonometric parallaxes of stars from the Gaia~DR2 catalog [6]. According to their estimates, the distances to molecular clouds are ultimately determined with an average error of about 5\%.

\begin{figure}[t]{ \begin{center}
  \includegraphics[width=0.95\textwidth]{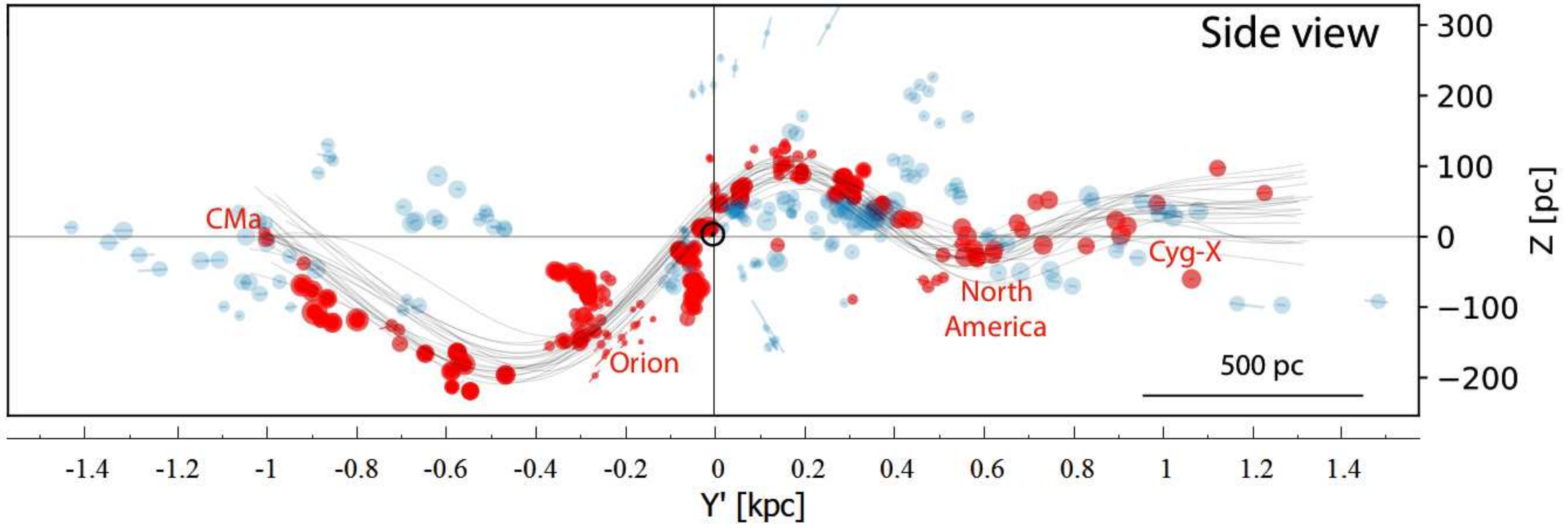}
  \caption{
Vertical coordinates of molecular clouds $z$ depending on the position on the $y'$ axis (this axis is located at an angle of $-30^\circ$ to the galactic axis $y$), the red circles indicate the clouds tracing the Radcliffe wave, the blue circles indicate field clouds, gray dotted lines are Radcliffe wave models. This is a drawing from Alves et al.~[1] to which we have added a more detailed $y'$ scale.}
 \label{f-Alves-00}\end{center}}\end{figure}

Fig.~\ref{f-Alves-00} shows the vertical coordinates of the molecular clouds identified by Alves et al.~[1] along the $y'$ axis, which is oriented at an angle of $30^\circ$ to the galactic axis $y $. The figure shows a number of Radcliffe wave models that were found in the work of these authors.

In this paper, we consider a heliocentric rectangular coordinate system $x,y,z$, in which the $x$ axis is directed from the Sun to the center of the Galaxy, the $y$ axis direction coincides with the direction of the Galaxy rotation, and the $z$ axis is directed to the northern Galactic pole, as well as the galactocentric rectangular coordinate system $X,Y,Z, $ in which the $X$ axis is directed from the center of the Galaxy to the Sun, the direction of the $Y$ axis coincides with the direction of rotation of the Galaxy, and the $Z$ axis is directed to the north galactic pole . Thus, in these two coordinate systems, only the directions of the $x$ and $X$ axes differ. In this case, the orientation of the Radcliffe wave with respect to the $y$ and $Y$ axes differs only in sign. For example, in the heliocentric coordinate system, the transition to the dashed axis $y'$ is carried out as follows:
\begin{equation}
 y'= y\cos{30^\circ}+x\sin{30^\circ}.
 \label{y'-30}
\end{equation}
In the light of what has been said, we can note that when citing the works of various authors, we do not distinguish between the designations $z$ and $Z$, as well as $y'$ and $Y'$. In addition, spatial velocities directed along the $x,y,z$ axes are usually denoted as $U,V,W$. At the same time, on the graphs from some of the cited works, the vertical velocities are designated as $V_z$.

The modeling of the Radcliffe wave by Alves et al.~[1] was done using a quadratic function in the coordinate space $x,y,z,$ given by three sets of ``control points'' $(x_0,y_0,z_0),$ $( x_1,y_1,z_1)$ and $(x_2,y_2,z_2).$ The undulating behavior about the wave center was described by a sinusoidal function about the $XY$ plane with a decaying period and amplitude:
  \begin{equation}
 \renewcommand{\arraystretch}{2.2}
 \label{Method-Alves}
 \begin{array}{lll}
 \Delta z(t)=
 A\times\exp\biggl[-\delta \biggl(\displaystyle{d(t)\over{\rm kpc}}\biggr)^2\biggr]\times
 \sin\biggl[\biggl(\displaystyle{2\pi d(t)\over P}\biggr)
 \biggl(1+{\displaystyle d(t)/d_{max}\over\gamma}\biggr)+\phi\biggr],
 \end{array}
 \end{equation}
where $d(t)=||(x,y,z)(t)-(x_0,y_0,z_0)||$~is the distance of a specific cloud from the beginning of the wave, specified by the parameter $t,$ $d_{ max}$~is distance from the end of the wave, $A$~ is wave amplitude, $P$~is wave period, $\phi$~ is wave phase, $\delta$ sets the decay rate amplitude, and  $\gamma$ specifies the decay rate of the period. As a result, Alves et al.~[1] found the following parameters of the model wave:
 \begin{equation}
 \renewcommand{\arraystretch}{1.0}
 \label{rez-Alves}
 \begin{array}{lll}
  \lambda=2.7\pm0.2~\hbox{kpc},\\
  A=160\pm30~\hbox{pc},\\
  \sigma_{\rm scatter}= 60\pm15~\hbox{pc},\\
  \hbox{Mass}\geq 3\times 10^6 M_\odot,
 \end{array}
 \end{equation}
where $\lambda$~is wavelength, $\sigma_{\rm scatter}$~is standard deviation of clouds from the model (wave radius). The given parameter errors in (\ref{rez-Alves}) correspond to the level of 95\% ($\pm 2\sigma$).

To date, about a dozen articles have already been published on the determination of the geometric and kinematic characteristics of the Radcliffe wave from data on various young objects, as well as hypotheses of its origin. The purpose of this paper is to review these publications.

\section{Confirmation of the existence of the Radcliffe wave}
 \subsection{Young stars, OSCs and dust clouds}
\begin{figure}[t]{ \begin{center}
  \includegraphics[width=0.95\textwidth]{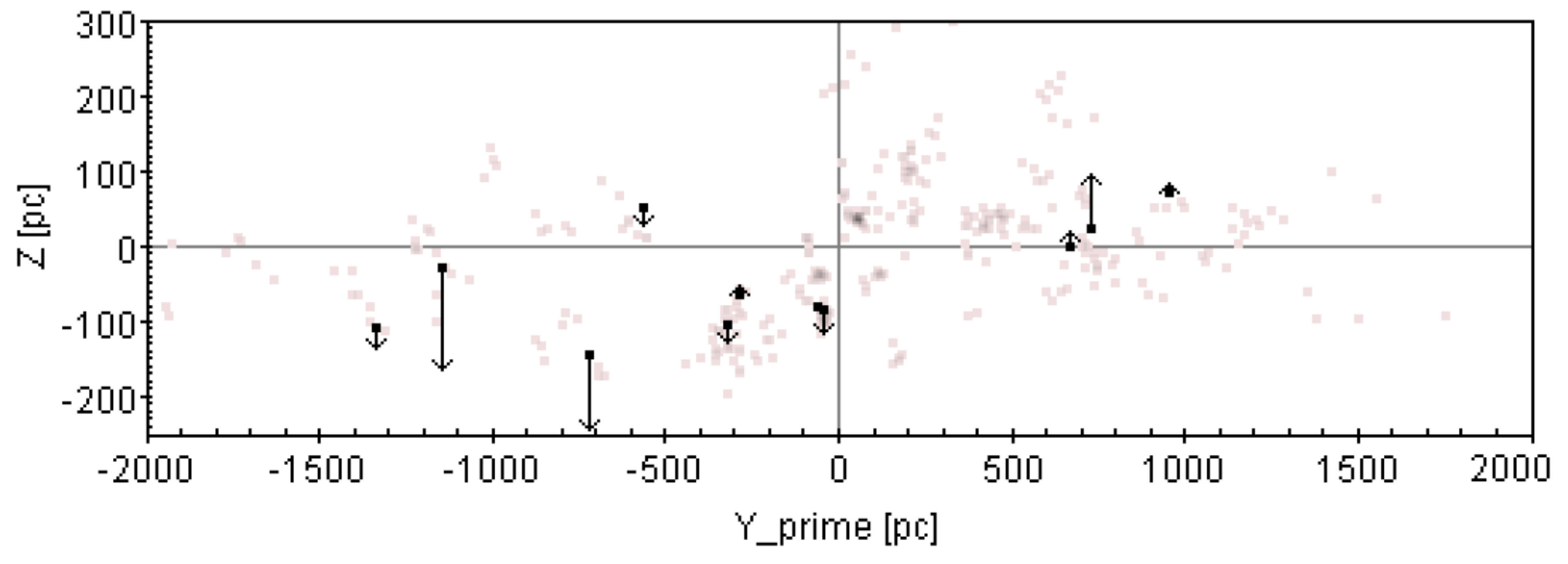}
  \caption{
Vertical coordinates $z$ of young OSCs versus distance $y'$. Vertical velocities $W$ of eleven ``OSC--MO'' pairs are shown by black arrows. The figure is taken from Donada, Figueras~[7].
 }
 \label{Donada-RW}\end{center}}\end{figure}

Donada and Figueras [7] analyzed a sample of very young OB stars and open star clusters (OSCs) younger than 30~Myr from the near-solar neighborhood with a radius of about 2~kpc. They developed criteria for cross matching these objects with identified cloud complexes belonging to the Radcliffe wave. Much attention is paid to the assessment of the quality of the used astrometric and photometric data, on the basis of which the estimates of distances to OB stars and OSCs were obtained. These authors made the first attempt to find a relationship between the structural and kinematic properties of the young stellar population associated with the Radcliffe wave and came to the following conclusions:
1) 13 OSCs have been identified, each of which is physically associated with molecular clouds (MOs) - probable members of the Radcliffe wave; 2)~compared to OSCs, single OB stars are less of an elongated structure that traces the Radcliffe wave; therefore, these authors focused their work on the analysis of OSCs; 3)~vertical motion of 11 ``OSC-MO'' pairs associated with the Radcliffe wave does not contradict a simple model of harmonic motion in the vertical direction, and 4)~trajectories of 13 OSCs integrated backward using the gravitational potential of the Galaxy do not assume an origin associated neither with a point nor with a straight line in the $XZ$ plane. As can be seen from Fig.~\ref{Donada-RW}, taken from the work of these authors, the relationship between vertical coordinates $z$ and vertical velocities $W$ of eleven OSCs belonging to the Radcliffe wave structure is quite clearly visible.

Swiggum et al.~[8] attempted to elucidate the spatial relationship between the Radcliffe wave and the Local Arm. For this, they used data on high-luminosity stars and young OSCs from the Gaia\,EDR3~[9] catalog in combination with 3D dust maps. This data set was examined in the context of color gradients observed in the spiral arms of other galaxies, where predictions from density wave theory and star formation models were applied to interpret the specific location of gas and dust clouds and OB stars. These authors concluded that the Radcliffe wave is a gas reservoir in the Local Arm, which is a laboratory for studying the formation of stars and molecular clouds in the Milky Way.

\begin{figure}[t]{ \begin{center}
  \includegraphics[width=0.55\textwidth]{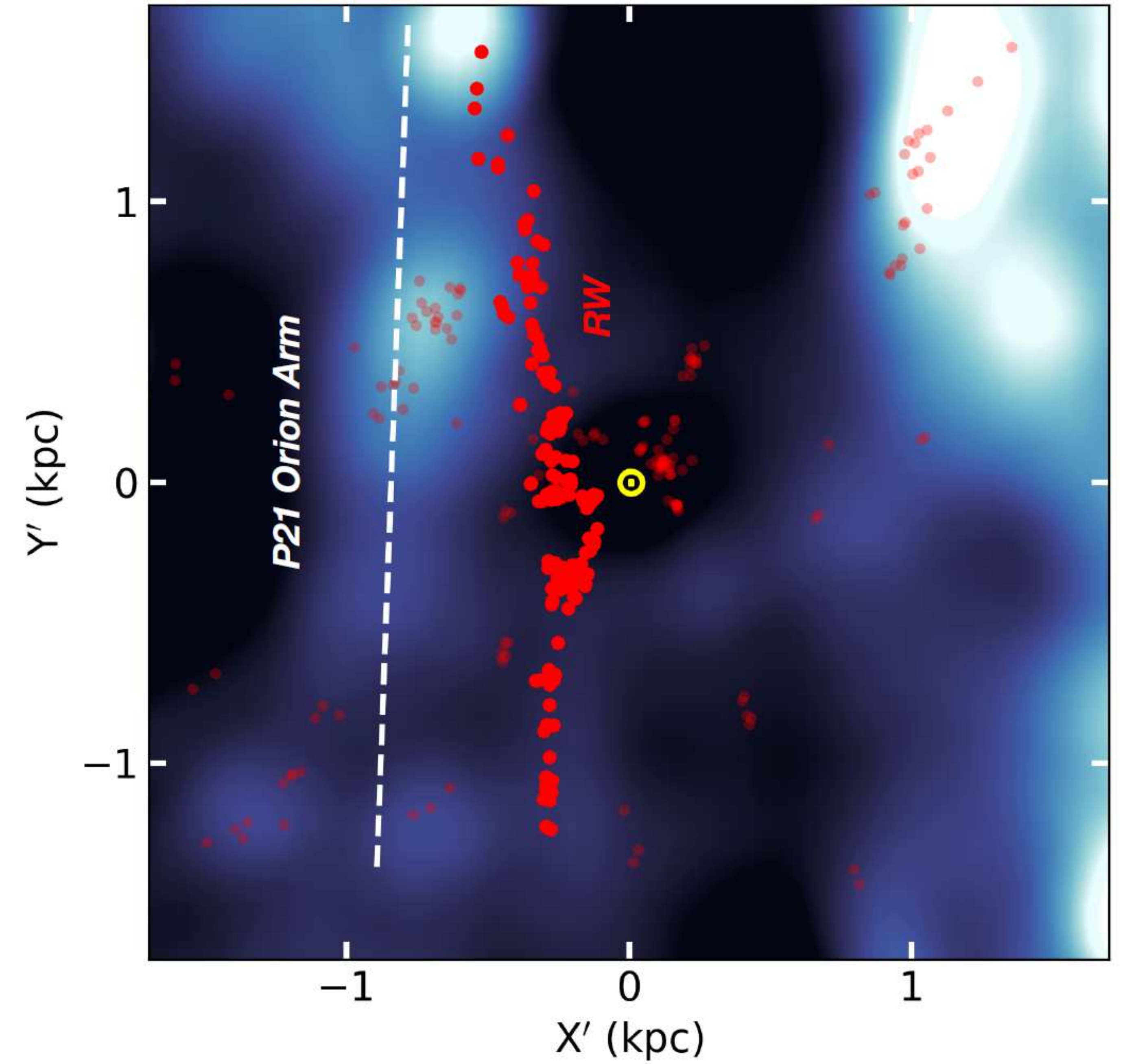}
  \caption{
In the rotated (by about 30$^\circ$ relative to the $XY$ coordinate system)  $X'Y'$ coordinate system, the distribution of star density is given in black-blue-white shading, the distribution of star-forming regions is given in pale red circles, according to~[5] , thick red circles mark the clouds belonging to the Radcliffe wave (RW) by ~[1], the dotted white line corresponds to peaks of excess stellar density along the structure P21~[10], the position of the Sun is marked with a yellow symbol. The figure is taken from Swiggum et al.~[8].
 }
 \label{Swiggum-RW}\end{center}}\end{figure}

In Fig.~\ref{Swiggum-RW}, it can be seen that the chain of molecular clouds associated with the Radcliffe wave is located along the inner edge of the P21 structure. The structure of P21 was revealed in [10] from an analysis of the distribution of young stars, OSCs, and classical Cepheids. For these objects, data from the Gaia\,EDR3 catalog were used. Of course, this structure is part of the Local Arm. The only surprise is a rather large angle, about $30^\circ$, under which this structure is inclined to the $Y$ axis.

According to the definitions of various authors, the pitch angle of the four-armed spiral pattern in the Galaxy lies in the interval 10$^\circ$--15$^\circ$~[11--16].
Although the Local Arm is not a grand design spiral arm, it is closely related to the spiral structure of the Galaxy. The pitch angle values found from the objects of the Local Arm also lie in a rather narrow range of values, 9$^\circ$--16$^\circ$~[15--17].
Most likely, the P21 structure indicates a local deviation from the general orientation of the Local Arm.

\begin{figure}[t]{ \begin{center}
  \includegraphics[width=0.95\textwidth]{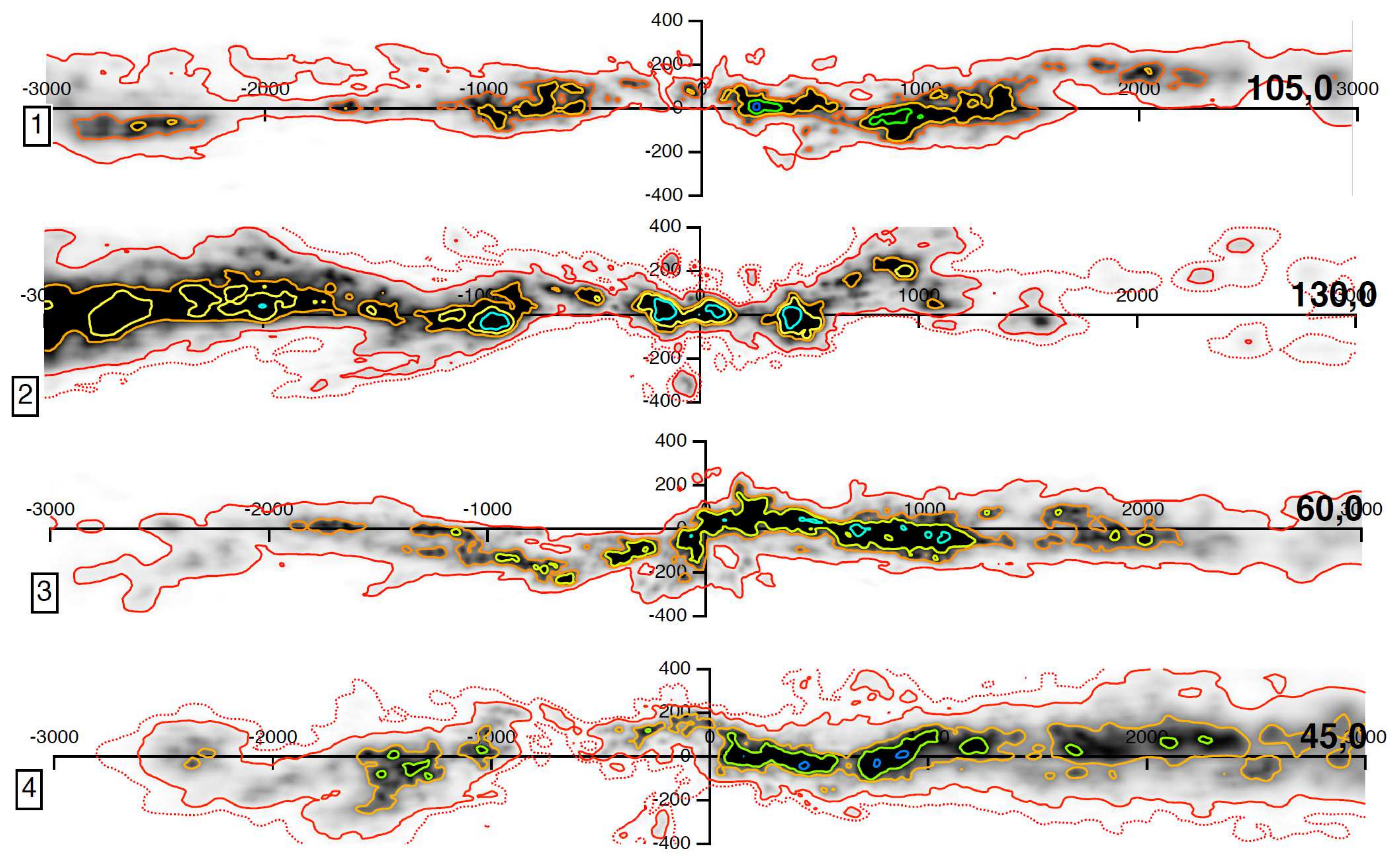}
  \caption{
Vertical distributions of interstellar dust in four sections passing near the Sun, where the Galactic longitude of the end of the section is indicated on the right, the section number is indicated on the left. The figure is taken from the work of Lallement et al.~[18].
 }
 \label{Lallement-RW}\end{center}}\end{figure}

Lallement et al.~[18] combined stellar photometric data from the Gaia\,EDR3 catalog with infrared measurements from the 2MASS~[19] catalog to construct a highly accurate 3D interstellar extinction map. Figure~\ref{Lallement-RW}, taken from~[18], shows the vertical distribution of interstellar dust in four horizontally narrow sections. As can be seen from this figure, wave-like deviations from the galactic plane in the vertical direction with an amplitude of up to 300~pc are observed in different directions. For us, the section number~3 is of greatest interest, oriented in the direction $l=60^\circ$ (thus, it passes at an angle $-30^\circ$ to the galactic axis $Y$), clearly showing the presence of the Radcliffe wave in the distribution of interstellar dust.

Thulasidharan et al.~[20] studied their vertical velocities from young stars located in the circumsolar region with a radius of 3~kpc. Three samples were analyzed: OB stars, stars from the upper part of the main sequence (these are mainly stars of spectral type A), and a sample of red giants. As a result, it was shown that the amplitude of vertical oscillations with a wavelength of about 2.5--2.7 kpc (along the $y'$ axis) depends on the age of the stellar population. The maximum amplitude of vertical velocities with a value of 3–4 km s$^{-1}$ is demonstrated by OB stars. In opinion of these authors, the reaction of the galactic disk to an external perturbation can serve as the main mechanism of the discovered such vertical oscillations.

Based on the study of the spatial distribution in the circumsolar neighborhood of OB stars with highly accurate distance estimates, Gonz\'alez et al. [41] identified an interesting structure, which they called the Cepheus spur. This structure is located inside the Local Arm, but it is located at a much larger angle to the $Y$ axis (at an angle of about $45^\circ$) than the arm itself. In the opinion of these authors, this structure is associated with the Radcliffe wave, since a wave-like character is observed in the distribution of vertical coordinates along the Cepheus spur (Fig.~7 in the work of Gonz\'alezet al. [41]). They also suggested that vertical oscillations in the galactic disk could be responsible for the recent enhanced star formation at the corresponding wave crests and troughs.

Li and Chen~[21] using data on a large number of young stars tracing the Radcliffe wave, found a relationship between their perturbed vertical positions and vertical velocities. For this purpose, low-mass stars that had not yet reached the main sequence stage were used. In this case, the vertical velocities of stars were calculated without using line-of-sight velocities (due to the absence of such measurements in the sample used). Therefore, the results of these authors should be considered as preliminary.

 \subsection{Method based on Fourier analysis}
To study the periodic structure in the coordinates and velocities of stars, Bobylev et al.~[22] proposed to use spectral analysis based on the standard Fourier transform of the original sequence $z(y')$:
\begin{equation}
 \renewcommand{\arraystretch}{2.4}
 F(z(y'))= \int z(y') e^{-j 2\pi y'/\lambda} =
 U(\lambda)+jV(\lambda)= A(\lambda) e^{j\varphi(\lambda)},
 \label{F}
\end{equation}
where
$A(\lambda)=\sqrt{U^2(\lambda)+V^2(\lambda)}$~is spectrum amplitude, and
  $\varphi(\lambda)=\arctan(V(\lambda)/U(\lambda))$~is spectrum phase.
A feature of this approach is the search for not just a monochromatic wave with a constant amplitude, but a wave that most accurately describes the initial data, the spectrum of which coincides with the main peak (lobe) of the calculated spectrum in the wavelength range from $\lambda_{min}$ to $\lambda_{ max}$ (within these boundaries, the spectrum gradually decreases starting from the maximum value, and outside~-- it begins to increase).

As a result, we have the desired smooth curve approximating the initial data, which is calculated by the formula of the inverse Fourier transform in the wavelength range we have defined:
\begin{equation}
 z(y')= 2k\int^{\lambda_{max}}_{\lambda_{min}} A(\lambda)\cos\biggl( {2\pi y'\over\lambda} + \varphi(\lambda) \biggr) d\lambda,\
 \label{Z}
\end{equation}
where $k$ is the coefficient calculated from the residual minimum condition.

\subsubsection{Fourier analysis of data from Alves et al.}
We first decided to test the Fourier analysis method using molecular cloud data from which Alves et al.~[1] discovered the Radcliffe wave.

In the combined sample, we used two data sets that were analyzed by Alves et al.~[1] and are publicly available: a)~data on dense clouds are taken from https://doi.org/10.7910/DVN/07L7YZ and b) ~data on more sparse structures are taken from https://doi.org/10.7910/DVN/K16GQX. Altogether, Alves et al.~[1] analyzed 380 structures with estimated distances to them. Moreover, random errors in estimating the distances to these molecular clouds, according to these authors, are about 5\%.

\begin{figure}[t]
{ \begin{center}
  \includegraphics[width=0.5\textwidth]{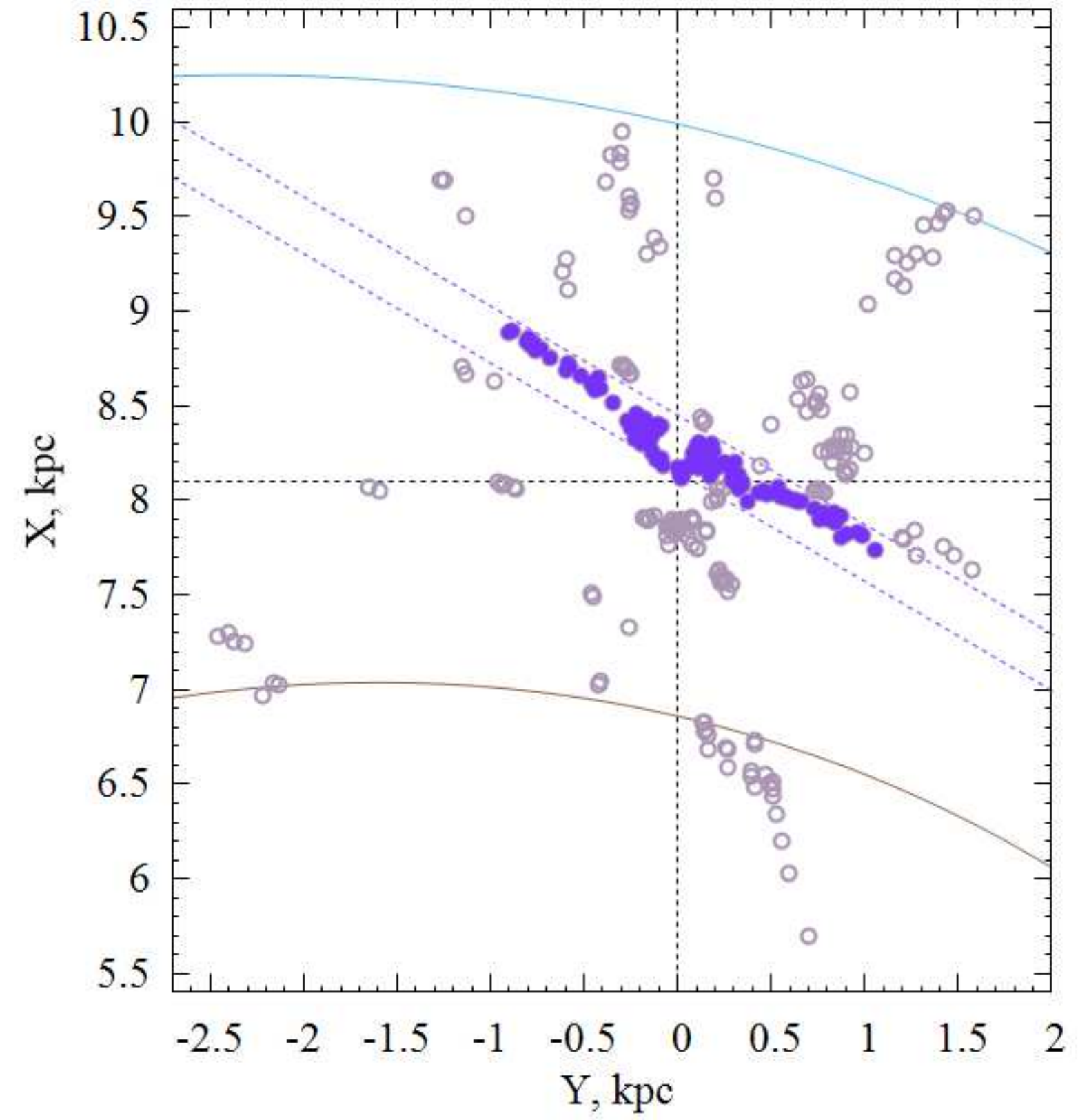}
  \caption{
Distribution of 380 molecular clouds from Alves et al.~[1] projected onto the galactic plane $XY$,~are given by gray circles, 189 clouds from a narrow zone passing at an angle of $-30^\circ$ to the $Y$ axis, are given by circles with a blue fill, there are marked two fragments of a four-armed spiral pattern with a pitch angle of $i=-13^\circ$.
  }
 \label{XY-Alves}
\end{center}}\end{figure}
\begin{figure}[t]{ \begin{center}
  \includegraphics[width=0.95\textwidth]{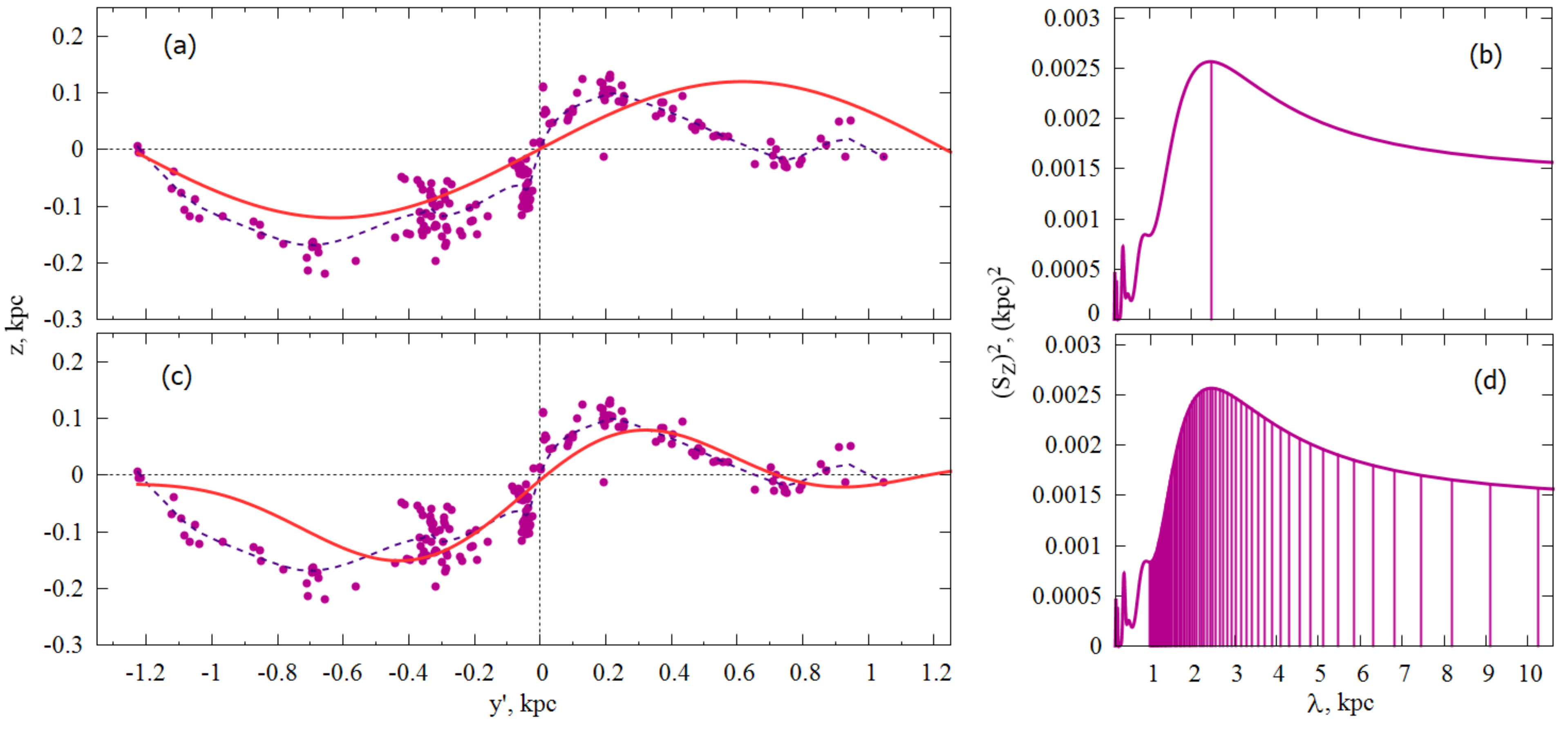}
  \caption{
Vertical coordinates $z$ of selected molecular clouds from Alves et al.~[1] versus distance $y'$~(a) and (c), their power spectra~(b) and (d), periodic heavy lines in graphs (a) and (c) reflect the corresponding results of the spectral analysis, and the dotted lines show the smoothed average values of the coordinates.
 }
 \label{Alves-poly-RW}\end{center}}\end{figure}

Figure~\ref{XY-Alves} shows the distribution of 380 molecular clouds projected onto the galactic $XY$ plane. To study the Radcliffe wave, we selected 189 of them located in a narrow zone inclined at an angle of $-30^\circ$ to the $Y$ axis, as shown in the figure. Thus, we have a sample very close to the one analyzed by Alves et al.~[1].
The figure shows fragments of a four-armed spiral pattern with a pitch angle $i=-13^\circ$ according to the work~[11].
Here this spiral pattern is constructed with the value of the distance from the Sun to the center of the Galaxy $R_0=8.1$~kpc. The value $R_0=8.1\pm0.1$~kpc was derived as a weighted average from a large number of modern individual estimates in the work~[39].

Then the positions of the selected molecular clouds were projected onto the $y'$ axis at an angle of $-30^\circ$ to the $y$ axis. In the coordinate system rotated in this way, a Fourier analysis of their positions is performed. Moreover, the analysis was carried out for two cases: a) for a monochromatic wave, when one frequency corresponding to the maximum of the spectrum is taken in the power spectrum, and b) for a polychromatic wave, when frequencies corresponding to the interval from $\lambda_{min}$ to $\lambda_{max}$ are taken in the power spectrum, according to the expression~(\ref{Z}). For each case, the value of the standard deviation $\sigma_z$ is calculated.

In the case of a monochromatic wave, the following estimates for the wave parameters were obtained:
  \begin{equation}
 \renewcommand{\arraystretch}{1.0}
 \label{rez-Z-mono}
 \begin{array}{lll}
  \lambda=2.5\pm0.1~\hbox{kpc},\\
  z_{max}=120\pm4~\hbox{pc},\\
  \sigma_z=60~\hbox{pc}.
 \end{array}
 \end{equation}
The parameter errors that we give here and below correspond to the 68\% ($\pm1\sigma$) level.
In the case of a polychromatic wave found:
 \begin{equation}
 \label{rez-Z-poly}
 \begin{array}{lll}
  \lambda=2.5\pm0.1~\hbox{kpc},\\
  z_{max}=150\pm4~\hbox{pc},\\
  \sigma_z=46~\hbox{pc}.
 \end{array}
 \end{equation}
Here, as well as in (\ref{rez-Z-mono}), the value of the wavelength $\lambda$ corresponds to the maximum of the power spectrum (Fig.~\ref{Alves-poly-RW}(d)). Note that in Fig.~\ref{Alves-poly-RW}(d) the first lobe of the spectrum is shaded. At the same time, the position of $\lambda_{min}$ is clearly visible, but the position of $\lambda_{max}$ is far beyond the picture.

In our method, we refused to designate the amplitude with the symbol $A$, since we indicate the maximum value of the wave $z_{max}$, which in this case is achieved at $y'=-0.43$~kpc. In the solution (\ref{rez-Z-poly}), we have a smaller dispersion value $\sigma_z.$ compared to (\ref{rez-Z-mono}). Thus, the polychromatic wave is in better agreement with the data. In addition, it also agrees better with the results of the analysis by Alves et al.~[1], which can be seen from a comparison of Fig.~\ref{f-Alves-00} and Fig.~\ref{Alves-poly-RW}, as well as from comparing found parameters (\ref{rez-Z-mono}) and (\ref{rez-Z-poly}) with (\ref{rez-Alves}). Note that Alves et al.~[1] also abandoned the use of a monochromatic wave in favor of a wave with damped amplitude and period (\ref{Method-Alves}).

Error estimates for the sought parameters were found using statistical Monte Carlo simulation based on 100 calculation cycles. With this number of cycles, the average values of the solutions practically coincide with the solutions obtained from the initial data without adding measurement errors. The measurement errors were added to the coordinates of the sources $x$, $y$, and $z$.

\begin{figure}[t]
{ \begin{center}
  \includegraphics[width=0.5\textwidth]{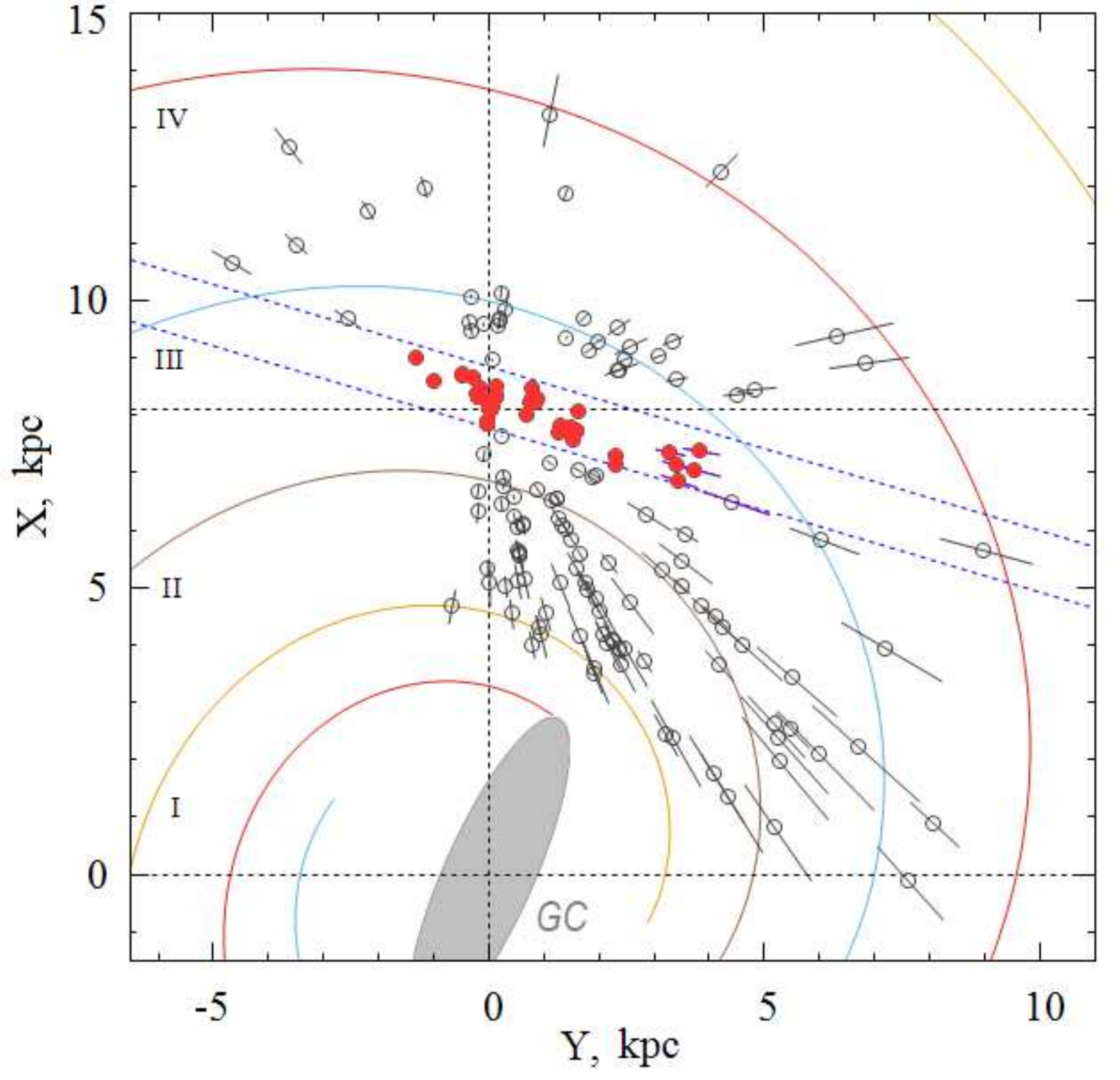}
  \caption{
Distribution of masers and radio stars with trigonometric parallax errors less than 15/%
in the projection on the galactic $XY$ plane. 68 masers and radio stars selected for analysis of the Radcliffe wave are marked with red circles,
there is shown a four-armed spiral pattern with a pitch angle of $i=-13^\circ$~[11], the central galactic bar is marked as well, GC~ is the center of the Galaxy. The figure is taken from the work of Bobylev et al.~[22].
  }
 \label{f-XY-masers}
\end{center}}
\end{figure}
\begin{figure}[t]
{ \begin{center}
  \includegraphics[width=0.9\textwidth]{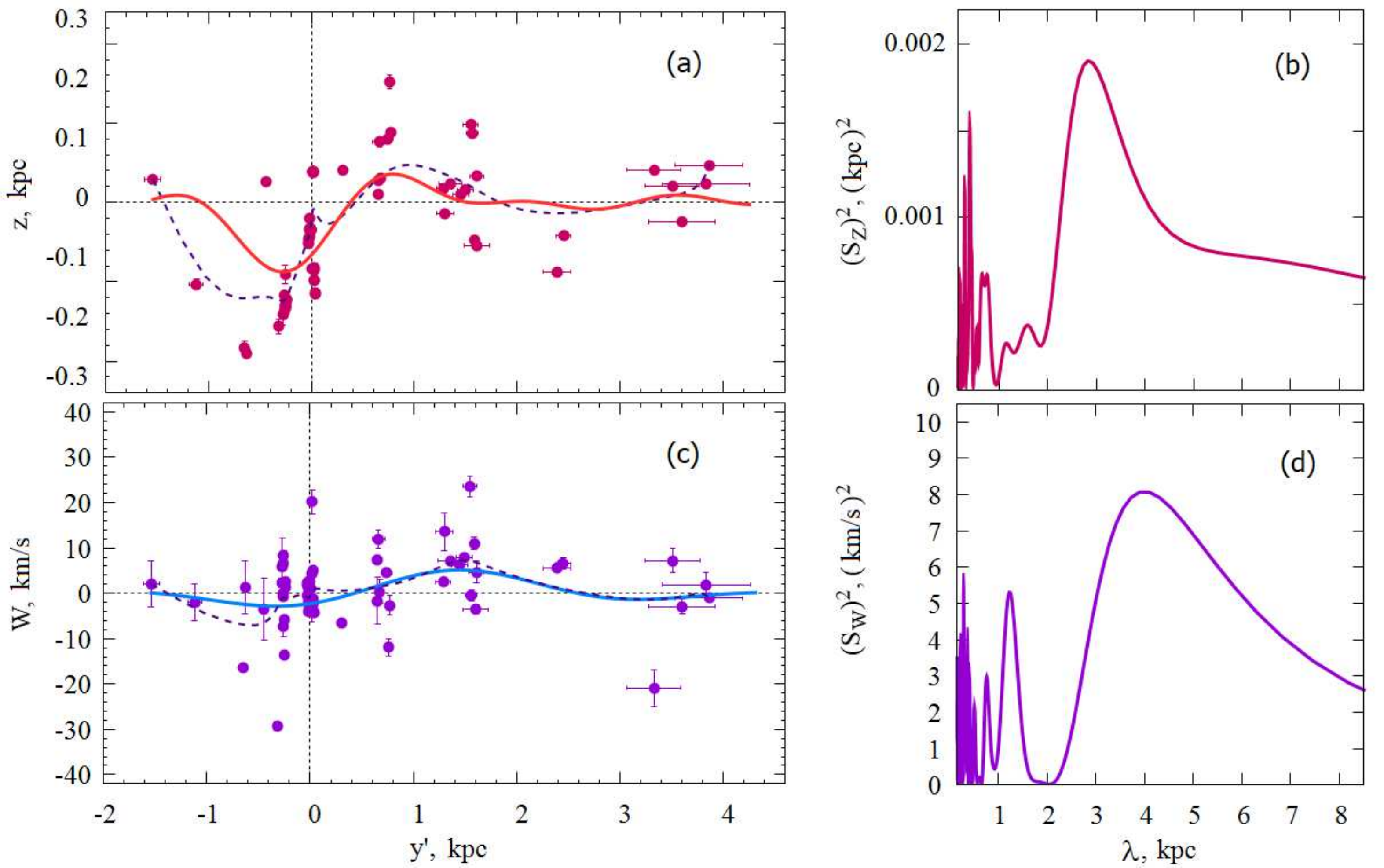}
  \caption{
Maser coordinates $z$ versus distances $y'$~(a) and their power spectrum~(b),
vertical velocities of masers $W$ versus distances $y'$~(c) and their power spectrum~(d). Periodic curves shown by solid thick lines reflect the results of spectral analysis, dotted lines show smoothed average values.
The figure is taken from the work of Bobylev et al.~[22].
}
 \label{f-spectr-masers}
\end{center}}
\end{figure}

\subsubsection{Fourier analysis of a sample of masers}
The sources of maser radiation are stars with extended gas-dust shells, in which the pumping effect occurs. Both very young stars and protostars of various masses, as well as old stars, such as Mirids, have the effect of maser radiation. To study the Radcliffe wave, Bobylev et al.~[22] used VLBI observations of only young objects that are closely associated with regions of active star formation.

It is important to note that astrometric VLBI observations of maser sources and radio stars are very accurate. So, the error in determining the trigonometric parallax is on average about 10 microseconds of  arc. This allows, in particular, to analyze the kinematics of masers with relative errors of distances of about 10\%, which are located up to the center of the galaxy.

The main data on the masers are contained in two large compilations: of  Reid et al. ~ [14] and Hirota et al. ~ [23]. Reid et al. ~ [14] gave information on 199 masers based on the results of VLBI observations by various authors on several radio frequencies as part of the Bessel project (The Bar and Spiral Structure Legacy Survey  {http://bessel.vlbi-astrometry.org} ).
In the work of Hirota et al. ~ [23] a catalog of 99 sources of maser radiation observed at a frequency of 22 ~ GHz according to the Vera (VLBI Exploration of Radio Astrometry  {http://verserver.naac.jp} ) is described.
Between the  samples of Reid et al. ~ [14] and Hirota et al. ~ [23] there is a large percentage of common  measurements. A number of new results of the determination of the parallax of the masers made after 2020 ~ [17, 24, 25] are also known.

In addition to the actual sources of maser radiation, the radio observations of which are carried out in narrow lines, we have radio stars in our list, the observations of which are made by the VLBI method in the continuum at a frequency of 8.4 ~ GHz ~ [26-29]. These are very young stars and protostars of type T \, Tauri, located mainly in the area of the Gould Belt and the Local Arm. Thus, these are objects that are closest by age to molecular clouds that were analyzed by Alves et al. ~ [1].

As can be seen from Fig.~\ref{f-XY-masers}, taken from~[22], there are not so many masers in the zone of interest to us to choose them in a narrow zone. Therefore, almost all sources located in the Local Spiral Arm were selected. A four-armed spiral pattern with a pitch angle $i=-13^\circ$ is given according to the work~[11]. Here this pattern is constructed with the value $R_0=8.1$~kpc, the following four spiral arms are numbered in Roman numerals: I~Scutum, II~--- Carina-Sagittarius, III~--- Perseus and IV~--- the Outer Arm. Red circles mark 68 sources selected for analysis. Due to the strong crowding of a number of nearby masers in the region of the associations of Orion, Taurus, or Scorpio-Centaurus, their projections merge into a point corresponding to each association in the figure. The two dotted blue lines inclined $-16^\circ$ to the $Y$ axis indicate the boundaries of the source selection area. The sampling zone is about 1.2 kpc wide. A constraint on the heliocentric distance of stars, $r<4$~kpc, was used as well.

The positions of the masers were projected onto the $y'.$ axis. And already in this rotated coordinate system, a spectral analysis of the positions and vertical velocities of the selected masers was carried out. As a result,from the analysis of the positions of the sources the following estimates were obtained for the maximum value of the $z$ coordinate ($z_{max}$, which is achieved at $y'=-0.28$~kpc) and the wavelength $\lambda$:
  \begin{equation}
 \label{sol-68-masers-Z}
 \begin{array}{lll}
  z_{max}= 87\pm4~\hbox{pc},\\
  \lambda=2.8\pm0.1~\hbox{kpc}.
 \end{array}
 \end{equation}
From the analysis of the vertical velocities of $W$ masers, we obtained an estimate of the maximum value of their perturbation velocity $W_{max}$ (which is achieved at $y'=1.4$~kpc) and the wavelength of these perturbations $\lambda$:
  \begin{equation}
 \label{sol-68-masers-W}
 \begin{array}{lll}
   W_{max}= 5.1\pm0.7~\hbox{km/s},\\
   \lambda= 3.9\pm1.6~\hbox{kpc}.
 \end{array}
 \end{equation}
The results of spectral analysis are shown in Fig.~\ref{f-spectr-masers}. The dotted lines in Fig.~\ref{f-spectr-masers}(a) and (c) show the smoothed averages of the data. The good agreement in the behavior of the solid and dotted lines in the circumsolar region indicates the reliability of the performed spectral analysis. We also note the agreement in the nature of the distribution along the wave of vertical velocities in Fig.~\ref{f-spectr-masers}(c) and Fig.~\ref{Donada-RW}.

\subsubsection{Fourier analysis of T Tauri stars}
Bobylev et al.~[22] used data on T Tauri stars to study the Radcliffe wave. The sample was based on the work of Marton et al.~[30], who selected young galactic stellar objects from a combination of orbital observations of space satellites~--- WISE~[31], Planck~[32] and Gaia~[33].
This database is called Gaia\,DR2$\times$AllWISE. It contains more than 100~million objects of various nature, which are divided into 4 main classes~--- young stellar objects (Young Stellar Objects, hereinafter YSO), main sequence stars, evolved stars and extragalactic objects. For each star, the probability of belonging to each of the four considered classes is determined. Probability estimates were found using the magnitudes G from the Gaia DR2~[6] catalog, the infrared photometric bands W1--W4 from the WISE catalog, and J,H,K from the 2MASS catalog. To decide how the source is related to the dust region, Marton et al.~[30] used the dust transparency index ($\tau$) for each object from the Planck map.

Parallaxes, proper motions and radial velocities of stars from the Gaia\,DR2$\times$AllWISE database were taken from the Gaia\,DR2 catalog in ~[34]. It turned out, however, that there are very few measured line-of-sight velocities for these stars. This does not allow one to calculate the full-fledged spatial velocities of stars. Therefore, only the spatial distribution of selected young stars was analyzed.

To select from the Gaia\,DR2$\times$AllWISE database the youngest stars that have not reached the main sequence stage, the following criteria were applied:
\begin{equation}
  \begin{array}{lll}
 {\rm LY}>0.95, ~~{\rm  SY}>0.98,\\
 {\rm LMS}<0.5, ~{\rm SMS}<0.5,\\
 {\rm  SE}<0.5, ~~~{\rm SEG}<0.5,
 \label{prob}
 \end{array}
\end{equation}
where
 SY~is the probability that the star is a YSO, found without using the W3 and W4 photometric bands from the WISE catalog,
 LMS~is the probability that the star is at the main sequence stage, found using all photometric bands from the WISE catalog,
 SMS~is the probability that the star is at the main sequence stage, found without using the W3 and W4 photometric bands from the WISE catalog,
 SE~is the probability that this is an evolving star found without involving the W3 and W4 photometric bands from the WISE catalog, and
 SEG~is the probability that this is an extragalactic source found without involving the W3 and W4 photometric bands from the WISE catalog.

It is known that the trigonometric parallaxes of stars from the Gaia\,DR2 catalog have a systematic shift with respect to the inertial coordinate system~[35].
In particular, it was shown in ~[35] that the value of such a correction is $\Delta\pi=-0.029$~mas. We used this value when calculating the distances $r$ to stars in terms of their parallaxes, $r=1/\pi_{true}$. Moreover, the use of the correction reduces the distance to the stars, because $\pi_{true}=\pi+0.029$.

To study the Radcliffe wave, stars on the galactic $XY$ plane were selected from a narrow zone located at an angle of $-25^\circ$ to the $Y$ axis. Stars with trigonometric parallax errors less than 10\% were used. In total, the sample included about 600 stars. We note that in the work of Thulasidharan et al.~[20], when analyzing three samples of young stars, the rotation to the $y'$ axis was also performed by $25^\circ$. The fact is that at a larger angle, there are very few distant stars in the sample .

\begin{figure}[t]
{ \begin{center}
  \includegraphics[width=0.5\textwidth]{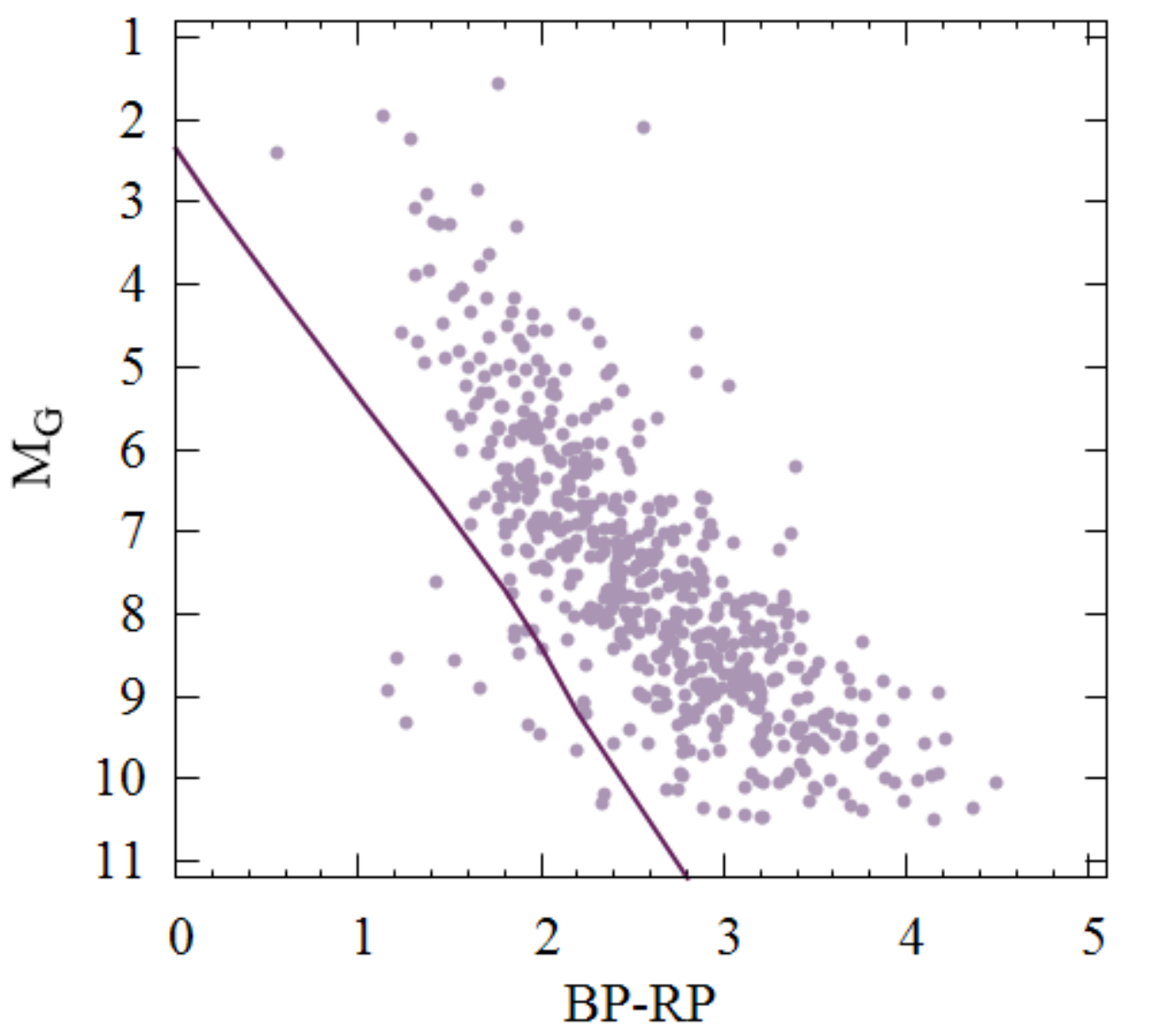}
  \caption{
Color index-absolute magnitude diagram plotted from stars in the Gaia\,DR2$\times$AllWISE database with relative parallax errors less than 10\%, the solid line marks the main sequence. The figure is taken from the work of Bobylev et al.~[22].  }
 \label{f-GR}
\end{center}}
\end{figure}

Figure~\ref{f-GR} shows a color index-absolute magnitude diagram constructed from a sample of stars from a zone passing at an angle of $-25^\circ$ to the $Y$ axis. The main sequence shown in the figure was drawn according to the work~[36]. Small details are of little interest to us, so the diagram is constructed without taking absorption into account. The main thing here is that the use of selection criteria~(\ref{prob}) makes it possible to select really very young stars that have not reached the main sequence stage.

\begin{figure}[t]{ \begin{center}
  \includegraphics[width=0.95\textwidth]{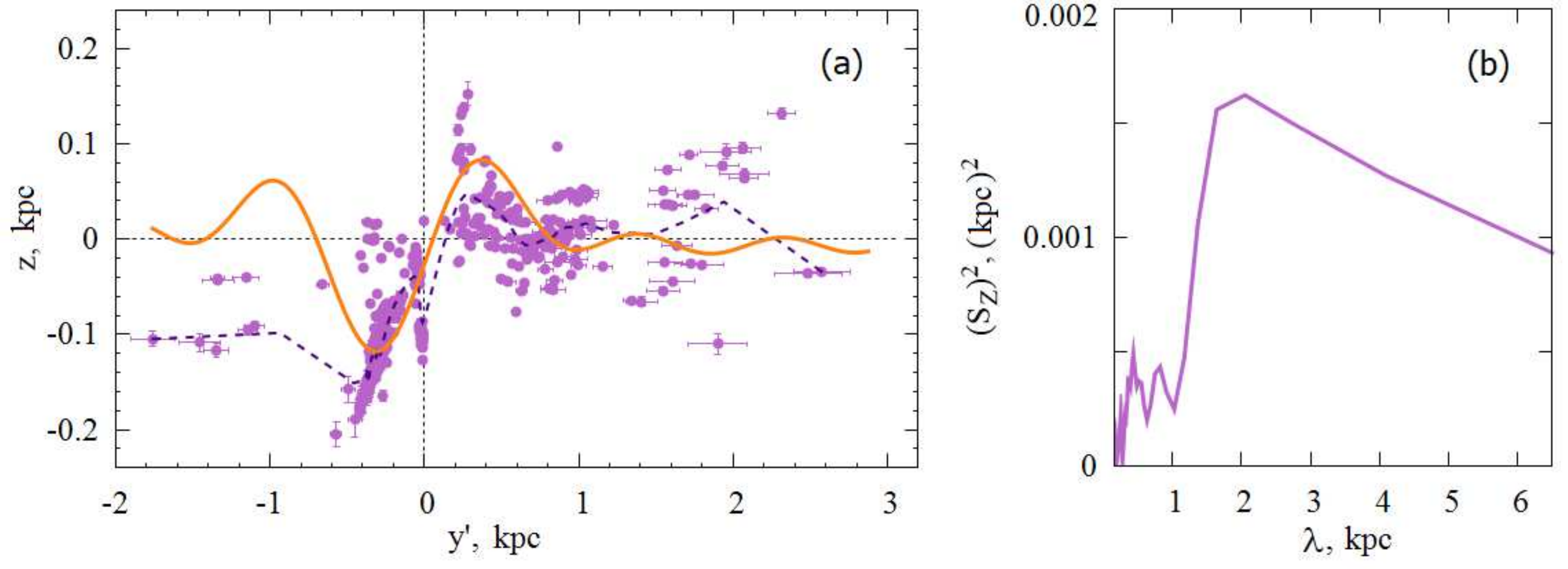}
  \caption{
YSO coordinates $z$ versus distance $y'$~(a) and their power spectrum~(b).
The curve line shows the results of the spectral analysis, the periodic thick line shows the result of the spectral analysis, the dotted line shows the smoothed average values of the coordinates. The figure is taken from the work of Bobylev et al.~[22].}
 \label{f-spectr-YSO}\end{center}}\end{figure}

Based on the Fourier analysis of this sample, the following estimates of the amplitude $z_{max}$ (which is achieved at $y'=-0.4$~kpc) and the wavelength $\lambda$ were obtained:
  \begin{equation}
 \label{sol-YSO-25}
 \begin{array}{lll}
  z_{max}=118\pm3~\hbox{pc},\\
  \lambda=2.0\pm0.1~\hbox{kpc}.
 \end{array}
 \end{equation}
These results are shown in Fig.~\ref{f-spectr-YSO}.


Let's note the work of Li and Chen~[21] in which, in order to study the Radcliffe wave, very young stars that have not reached the main sequence stage were also analyzed. In contrast to ours, here the basis for the classification of young stars was the earlier work by Marton et al.~[37]. The proper motions of stars from the Gaia\,DR2 catalog were used to analyze the kinematics. In this case, the vertical velocities of the stars were calculated without using line-of-sight velocities. An original method was used to search for the parameters of the observed wave in the positions and vertical velocities of the stars. These authors concluded that the vertical positions and velocities of the stars demonstrate almost the same periodicity with a wavelength $\lambda$ of about 1.5~kpc, both oscillations have a damped character, the amplitude of oscillations relative to the average plane of the Milky Way disk is $z_{max}=130\pm20 $~pc. We can see that there is good agreement in the estimates of $\lambda$ and $z_{max}$ found by Lee and Chen~[21] and by us in the solution (\ref{sol-YSO-25}).

\subsubsection{Vertical velocities in the Radcliffe wave}
Let's pay attention to Fig.~\ref{Donada-RW} and Fig.~\ref{f-spectr-masers}, from which we can see that the oscillations of the vertical coordinates and vertical velocities of the stars are almost synchronous. A similar behavior of the coordinates and vertical velocities of young stars was obtained by Thulasidharan et al.~[20].

\begin{figure}[t]{ \begin{center}
  \includegraphics[width=0.8\textwidth]{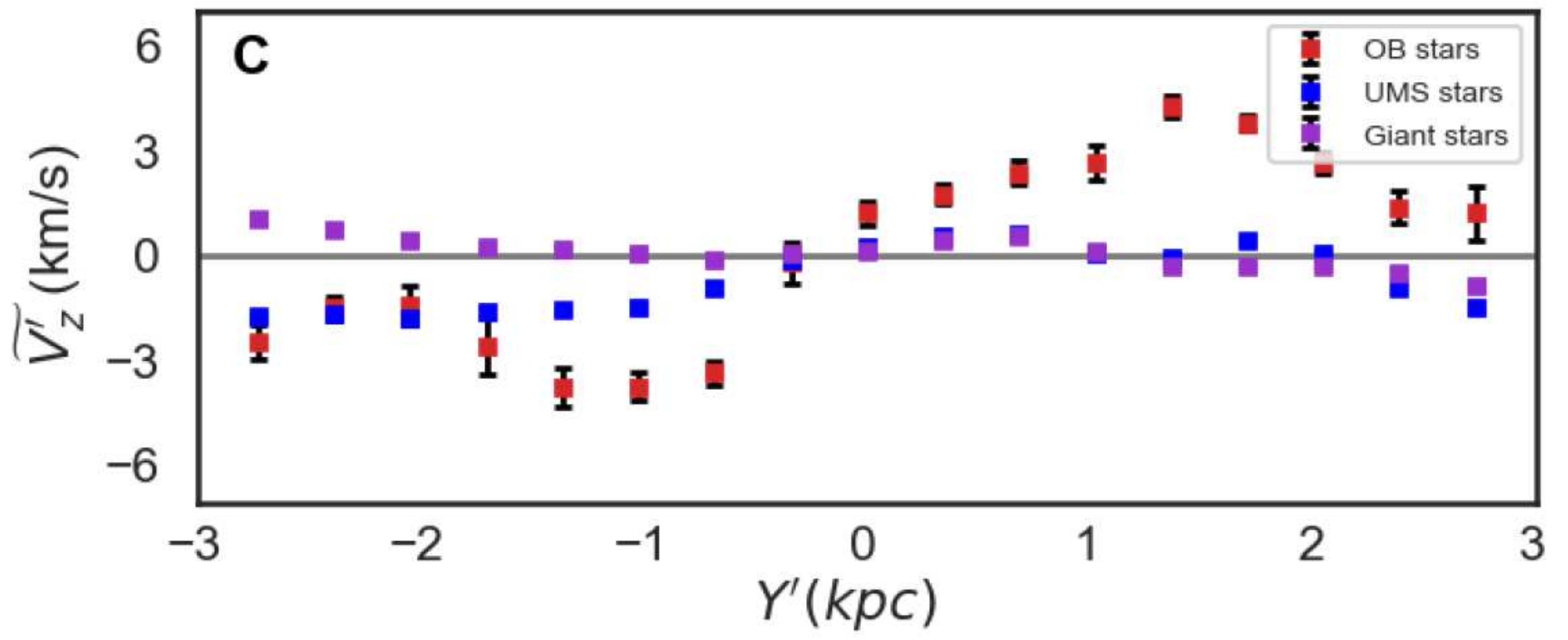}
  \caption{
Vertical velocities of three samples of young stars depending on the distance $y'$, the position of the Sun here approximately corresponds to the value $y'=0$~kpc. The figure is taken from Thulasidharan et al.~[20]. }
 \label{f-Tula-W}\end{center}}\end{figure}

An analysis of the velocities of three samples of young stars performed by Thulasidharan et al.~[20] is shown in Fig.~\ref{f-Tula-W}. The ``C'' symbol in the upper left corner of this figure means that the stars belong to a narrow band oriented at an angle of $25^\circ$ to the $Y$ axis. The vertical velocities of the samples are given in different colors: the velocities of OB stars are given in red, the velocities of stars from the upper main sequence (UMS) are given in blue, and the velocities of giants are given in purple. Note the good agreement between the wave of vertical velocities of OB stars and masers (Fig.~\ref{f-spectr-masers}).

\begin{figure}[t]{\begin{center}
 \includegraphics[width=0.8\textwidth]{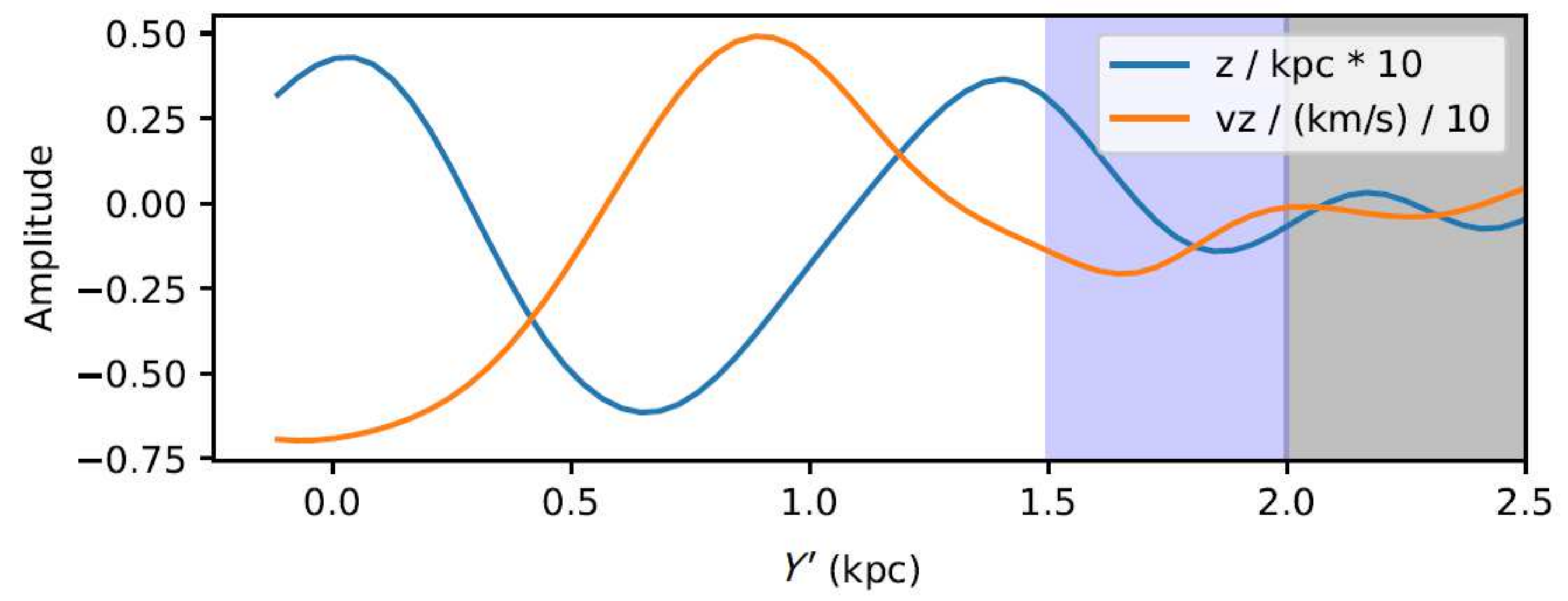}
  \caption{
Behavior of vertical coordinates (blue line) and vertical velocities (orange line) of T\, Tauri type stars depending on the distance $y'$, the position of the Sun here approximately corresponds to the value $y'=1.0$~kpc. The figure is taken from Li and Chen~[21].}
 \label{f-Li-W} \end{center}} \end{figure}

But Li, Chen~[21] from the analysis of the proper motions of stars that have not reached the main sequence stage, found a phase difference with a value of about $2\pi/3$ between the wave of vertical coordinates and the wave of vertical velocities. This is shown in Fig.~\ref{f-Li-W}, which we took from the work of these authors with a slight change in the designation of the abscissa axis. In this case, the amplitude of the vertical velocities of disturbances, $W_{max}=7\pm0.6$~km s$^{-1}$, is in good agreement with the estimates of other authors. Once again, we note that these authors calculated the vertical velocities of stars without using line-of-sight velocities. These stars are very faint, so far it has not been possible to obtain their spectra. Therefore, mass measurements of the line-of-sight velocities of these stars are not yet available. Thus, the analysis by Li, Chen~[21] of the vertical velocities of young stars must be treated with caution.

Thus, we can say with great confidence that the oscillations of the vertical positions and vertical velocities of stars in the Radcliffe wave occur synchronously.

We note the work of Tu et al.~[42], who performed a three-dimensional analysis of the positions and velocities of about 1100 young stars that have not reached the main sequence stage associated with the Radcliffe wave. In contrast to [21], to calculate the spatial velocities of stars, Tu et al.~[42] used stars with measured line-of-sight velocities (from the Gaia\,DR2 catalogue), proper motions, and parallaxes. The basis of the work was the catalog of young stars from the work of Zari et al.~[43], where all the stars are located no further than 500 pc from the Sun. Tu et al.~[42] found good agreement between the behavior of gas and dust along the Radcliffe wave known from Alves et al.~[1] and the behavior of the vertical coordinates of young stars. At the same time, these authors note that the amplitude of vertical perturbations found from a sample of young stars is slightly less than that for gas and dust.

These authors carried out a rather complicated analysis of the velocities of selected young stars~--- galactic orbits were constructed in several potentials. Then the positions and velocities of the stars were expressed in terms of angle-action, and then the tendency to change the vertical angle of the star ($\Omega_z$) depending on its position along the Radcliffe wave (we have along the $y'$ axis) was considered. The main conclusion of Tu et al.~[42] on the kinematics of young stars is that they did not find significant changes in vertical velocities depending on the position of the stars along the Radcliffe wave. They attribute this to the quality of line-of-sight velocities of the stars under consideration and express the hope that more accurate mass line-of-sight velocities of faint stars will appear (for example, in the Gaia\,DR3 catalog).

\section{Hypotheses of the origin of the Radcliffe wave}
According to Fleck~[38], the origin of the Radcliffe wave could be caused by the Kelvin--Helmholtz instability, which appeared at the interface between the galactic disk and the halo rotating at different velocities. Simplifying, we can say that the effect is similar to the waves from the wind on the river. Thus, the critique of Fleck's approach is that if the effect works, Radcliffe-type waves should be observed throughout the galactic disk.

Using data on young stars as indicators of the Radcliffe wave, Thulasidharan et al. [20] explored the possibility that this oscillation is a part of a larger vertical mode. According to these authors, there is a kinematic wave in the Galaxy with an oscillation amplitude that depends on the age of the stellar population.
Based on N-body simulation, the gravitational effect on the galactic disk of a dwarf satellite galaxy of the Milky Way type of the well-known dwarf galaxy in Sagittarius with a mass of about $2\times10^{10} M_\odot$ is studied. These authors found that the impact of such an impactor induces a kinematic wave propagating in a direction radial from the center of the Galaxy with an amplitude of vertical oscillations of 4-5~km s$^{-1}$ in the region of the Sun. However, the pitch angle of the model wave was obtained too large compared to that required to explain the appearance of the Radcliffe wave. The authors postponed their final conclusion until more advanced observational data became available.

Note that a globular cluster can also act as a striker. For example, Bobylev and Bajkova~[40] showed that the globular cluster $\omega$~Cen could have caused the emergence of the Gould Belt. True, the passage of this globular cluster through the galactic disk, according to the estimates obtained, should have taken place about 90 million years ago. To evaluate the relationship of this effect with the Radcliffe wave, it is necessary to know the age of the wave. The manifestation of the Radcliffe wave is observed in the distribution of OB stars [20]. Therefore, we can say that the age of the Radcliffe wave is at least 40–50 Myr. In principle, this is comparable to the age of the Gould Belt, $\sim60$ Myr.

At present, a number of phenomena of various nature are known, demonstrating the presence of vertical oscillations in the disk of the Galaxy. For example, large-scale disk curvature [44] observed in the distribution and kinematics of gas and dust clouds, as well as stars [45–49], is well established. The probability of a connection between the Radcliffe wave and this phenomenon is negligible due to the characteristic scale.

There are possible vertical oscillations associated with perturbations from the Galactic spiral density wave [50, 51]. The characteristic wavelength of such oscillations (2–3 kpc) is close to that found for the Radcliffe wave. However, the orientation of the Radcliffe wave (tilt to the $y$ axis) is very different from the pitch angle of the spiral pattern of the Galaxy ($10-15^\circ$). Therefore, the connection of the Radcliffe wave with perturbations from the helical density wave, although not excluded, still seems unlikely.

The papers [52, 53] describe periodic perturbations in the density and velocities of stars that are asymmetric in the vertical direction, i.e., which are asymmetric for the northern and southern galactic hemispheres. These studies used main-sequence stars close to the Sun. Thus, stars with ages of hundreds of million years were involved. Such an imbalance in the galactic disk is usually associated with some external influence on the disk of the Galaxy (the fall of a garlic satellite galaxy, a dark matter bunch, a massive high-speed gas cloud, a globular cluster, etc.).

There are a number of questions about the nature of the Radcliffe wave that have not yet been answered. First, there is no complete certainty that this is a wave. Secondly, how do you know that the wave is damped? If it moves in the direction of rotation of the Galaxy, then yes, the wave is damped. And if against the rotation of the Galaxy, then the wave will no longer be damped.

As a result, we can conclude that the Radcliffe wave has unique characteristics. Such characteristics are difficult to explain. Therefore, the Radcliffe wave remains mysterious for now.

\section{Conclusion}
In 2020, Alves et al.~[1] discovered the Radcliffe wave~--- a wave of damped vertical oscillations from molecular cloud data. The characteristic scale of the wave is about 2.5~kpc, and the maximum amplitude is 160 pc. This paper reviews a number of publications that confirm the presence of the Radcliffe wave both in the positions and in the vertical velocities of various young stars. We have tried to note all the publications available to date, in which the term Radcliffe wave appears either in the title or in the keywords.

Donada and Figueras [7] analyzed a sample of very young OB stars and OSCs from the solar neighborhood with a radius of about 2~kpc. These authors apparently discovered for the first time the relationship between the vertical coordinates and vertical velocities of young objects belonging to the structure of the Radcliffe wave. In this case, vertical velocities of only 11 ~ OSC were used.

In the work of Lallement et al.~[18], based on modern photometric data on stars, high-precision three-dimensional maps of interstellar extinction were constructed and the presence of a Radcliffe wave in the distribution of interstellar dust was shown.

Thulasidharan et al.~[20] analyzed a) OB stars, b) upper main sequence stars, and c) red giants. It is shown that the amplitude of vertical oscillations depends on the age of the stellar population. The relationship between the perturbed vertical positions and vertical velocities of young objects has been confirmed. The maximum amplitude of vertical velocities with a value of 3–4 km s$^{-1}$ is demonstrated by OB stars. Moreover, the vertical velocities of a large number of stars have already been studied here.

An analysis of young T\, Tauri type stars by Li, Chen~[21], and Bobylev et al.~[22] showed the presence of a Radcliffe wave at the positions of these stars.

In addition to T\ and Tauri type stars, Bobylev et al.~[22] studied a sample of 68 maser sources and radio stars located in the Local Arm with high-precision VLBI measurements of their trigonometric parallaxes and proper motions. The analysis of their positions and velocities is based on the Fourier analysis. The use of this method makes it possible to inscribe both a monochromatic and a polychromatic wave into the observational data. Specifically for this work, Fourier analysis was applied to the molecular cloud data that were used to detect the Radcliffe wave. The method has shown excellent results.

When analyzing the proper motions of stars that did not reach the main sequence stage, Li, Chen~[21] obtained a paradoxical result~--- they found a phase difference with a value of about $2\pi/3$ between the wave of vertical coordinates and the wave of vertical velocities. In this paper, we have actually discussed the results of the analysis of the vertical velocities of stars associated with the Radcliffe wave obtained by Donada, Figueras~[7], Thulasidharan et al.~[20], and Bobylev et al.~[22]. As a result, we concluded that the fluctuations in the vertical positions and vertical velocities of stars in the Radcliffe wave occur synchronously.

To explain the nature of the occurrence of the Radcliffe wave, two hypotheses have been put forward to date. Fleck~[38] proposes to relate the origin of the Radcliffe wave to the Kelvin--Helmholtz instability. The majority of researchers, however, adhere to the assumption of an external gravitational effect on the galactic disk of an impactor such as a dwarf satellite galaxy of the Milky Way.

\newpage

\bigskip\medskip{REFERENCES}\medskip {\small

\noindent
1. J. Alves, C. Zucker, A.A. Goodman, et al., Nature {\bf 578}, 237, 2020.

\noindent
2. V.V. Bobylev, Astrophysics {\bf 57}, 583, 2014.

\noindent
3. A. Blaauw, Koninkl. Ned. Akad. Wetenschap. {\bf 74}, No.~4, 1965.

\noindent
4. C. Zucker, J.S. Speagle, E.F. Schlafly, et al., Astrophys. J. {\bf 879}, 125, 2019.

\noindent
5. C. Zucker, J.S. Speagle, E.F. Schlafly, et al., Astron. Astrophys. {\bf 633}, 51, 2020.

\noindent
6. Gaia Collab. (A.G.A. Brown, A. Vallenari, T. Prusti, et al.),
  Astron. Astrophys. {\bf 616}, 1, 2018. 

\noindent
7. J. Donada, F. Figueras, arXiv: 2111.04685, 2021.

\noindent
8. C. Swiggum, J. Alves, E. D'Onghia, R.A. Benjamin, et al., arXiv: 2204.06003, 2022.

\noindent
9. Gaia Collab. (A.G.A. Brown, A. Vallenari, T. Prusti, et al.),
 Astron. Astrophys. {\bf 649}, 1, 2021. 

\noindent
10. E. Poggio, R. Drimmel, T. Cantat-Gaudin, et al., Astron. Astrophys.
Suppl. Ser. {\bf 651}, 104, 2021.

\noindent
11. V.V. Bobylev and A.T. Bajkova, MNRAS {\bf 437}, 1549, 2014. 

\noindent
12. J.P. Vall\'ee, Astron. J. {\bf 135}, 1301, 2008.

\noindent
13. J.P. Vall\'ee, New Astron. Review {\bf 79}, 49, 2017.

\noindent
14. M.J. Reid, N. Dame, K.M. Menten, et al., Astrophys. J. {\bf 885}, 131, 2019. 

\noindent
15. C.J. Hao, Y. Xu, L.G. Hou, et al., Astron. Astrophys. {\bf 652}, 102, 2021.

\noindent
16. L. Martinez-Medina, A. P\'erez-Villegas, and A. Peimbert, MNRAS {\bf 512}, 1574, 2022.

\noindent
17. Y. Xu, L.G. Hou, S. Bian, et al., Astron. Astrophys. {\bf 645}, L8, 2021.

\noindent
18. R. Lallement, J.L. Vergely, C. Babusiaux, et al., Astron. Astrophys. {\bf 661}, 147, 2022.

\noindent
19. M.F. Skrutskie, R.M. Cutri, R. Stiening, et al., Astron. J. {\bf 131}, 1163, 2006.

\noindent
20. L. Thulasidharan, E. D'Onghia, E. Poggio, et al., Astron. Astrophys. {\bf 660}, 12, 2022.

\noindent
21. G.-X. Li and B.-Q. Chen, arXiv: 2205.03218, 2022.

\noindent
22. V.V. Bobylev, A.T. Bajkova, and Yu.N. Mishurov, Astron. Lett. {\bf 48}, 999, 2022.

 \noindent
23. VERA Collab. (T. Hirota, T. Nagayama, M. Honma, et al.), PASJ {\bf 70}, 51, 2020.

 \noindent
24. N. Sakai, H. Nakanishi, K. Kurahara, et al., PASJ {\bf 74}, 209, 2022.

 \noindent
25. S.B. Bian, Y. Xu, J.J. Li, et al., Astron. J. {\bf 163}, 54, 2022.

 \noindent
26. R.M. Torres, L. Loinard,  A.J. Mioduszewski, et al., Astrophys. J. {\bf 671}, 1813, 2007.

 \noindent
27. S. Dzib, L. Loinard, L.F. Rodriguez, et al., Astrophys. J. {\bf 733}, 71, 2011.

 \noindent
28. G.N. Ortiz-Le\'on, L. Loinard, S.A. Dzib, et al., Astrophys. J. {\bf 865}, 73, 2018.

 \noindent
29. P.A.B. Galli, L. Loinard, G.N. Ortiz-L\'eon, et al.,
 Astrophys. J. {\bf 859}, 33, 2018.

 \noindent
30. G. Marton, P. \'Abrah\'am, E. Szegedi-Elek, et al., MNRAS {\bf 487}, 2522, 2019.

 \noindent
31. E.L. Wright, P.R.M. Eisenhardt, A.K. Mainzer, et al., Astroph. J. {\bf 140}, 1868, 2010.

 \noindent
32. Planck Collab. (R. Adam, P.A.R. Ade, N. Aghanim, et al.),  Astron. Astrophys. {\bf 594}, 10, 2016.

 \noindent
33. Gaia Collab. (T. Prusti, J.H.J. de Bruijne, A.G.A. Brown, et al.), Astron. Astrophys. {\bf 595}, A1, 2016. 

 \noindent
34. O.I. Krisanova, V.V. Bobylev, and A.T. Bajkova, Astron. Lett. {\bf 46}, 370, 2020.

 \noindent
35. Gaia Collab. (L. Lindegren, J. Hernandez, A. Bombrun, et al.),
 Astron. Astrophys. {\bf 616}, 2, 2018.

 \noindent
36. E. Zari, H. Hashemi, A.G.A. Brown, et al., Astron. and Astrophys. {\bf 620}, 172, 2018.

\noindent
37. G. Marton, L.V. T\'oth, R. Paladini, et al., MNRAS {\bf 458}, 3479, 2016.

\noindent
38. R. Fleck, Nature {\bf 583}, 24, 2020.

\noindent
39. V.V. Bobylev, A.T. Bajkova, Astron. Rep. {\bf 65}, 498, 2021. 

\noindent
40. V.V. Bobylev, A.T. Bajkova, Astron. Rep. {\bf 62}, 557, 2018. 

\noindent
41. M.P. Gonz\'alez, J. Maiz Apell\'aniz, R.H. Barb\'a and B. C. Reed,
 MNRAS {\bf 404}, 2967, 2021.

\noindent
42. A.J. Tu, C. Zucker, J.S. Speagle, et al., arXiv: 2208.06469, 2022.

\noindent
43. E. Zari, H. Hashemi, A.G.A. Brown, et al., Astron. Astrophys. {\bf 620}, 172, 2018.

\noindent
44. G. Westerhout, Bull. Astron. Inst. Netherlands {\bf 13}, 201, 1957. 

\noindent
45. D. Russeil, Astron. Astrophys. {\bf 397}, 133, 2003.

\noindent
46. I. Yusifov, astro-ph/0405517, 2004.

\noindent
47. R. Drimmel, R.L. Smart, and M.G. Lattanzi, Astron. Astrophys. {\bf 354}, 67, 2000.

\noindent
48. M. L\'opez-Corredoira, A. Cabrera-Lavers, F. Garz\'on, and P.L. Hammersley,
Astron. Astrophys. {\bf 394}, 883, 2002.

\noindent
49. L.N. Berdnikov, Astron. Lett. {\bf 13}, 45, 1987.

\noindent
50. Yu.N. Mishurov, Astron. Rep. {\bf 50}, 12, 2006.

\noindent
51. V.V. Bobylev and A.T. Bajkova, MNRAS {\bf 447}, L50, 2015. 

\noindent
52. L. M. Widrow, S. Gardner, B. Yanny, S. Dodelson, and H.-Y. Chen,
 Astrophys. J. Lett. {\bf 750}, L41, 2012.

\noindent
53. M. Bennett, J. Bovy,  MNRAS {\bf 482}, 1417, 2018.

}
\end{document}